\documentclass{ws-p9-75x6-50}

\newcommand{\tr}{{\mathrm{tr}}}
\newcommand{\str}{{\mathrm{str}}}

\newcommand{\one}{\hbox{ 1\kern-.8mm l}}

\newcommand{\bear}{\begin{array}{l}}
\newcommand{\eear}{\end{array}}
\newcommand{\ds}{\displaystyle}

\newcommand{\ie}{{\it i.e.}\ }
\newcommand{\eg}{{\it e.g.}\ }
\newcommand{\secti}{sec.\ }
\newcommand{\csec}{Sec.\ }

\def\rtil{\tilde{r}}
\def\rtilo{\tilde{r}_0}
\def\rhat{\hat{r}}
\def\A{{\cal A}}
\def\C{{\cal C}}
\def\S{{\cal S}}
\def\rhato{\hat{r}_0}
\def\dga{${\cal D}_{\Gamma}$}
\def\wick#1#2{\hspace{-#1mm}\raisebox{-2ex}{\rule{0.02mm}{2mm}\rule{#2mm}{0.02mm}\rule{0.02mm}{2mm}}\hspace{#1mm}\hspace{-#2mm}}
\def\eq#1{eq.~(\ref{#1})}

\def\Box#1{\mathop{\mkern0.5\thinmuskip
               \vbox{\hrule
                  \hbox{\vrule
                        \hskip#1
                        \vrule height#1 width 0pt
                        \vrule} %
                  \hrule}      
               \mkern0.5\thinmuskip}}
\def\hepth#1 {{\tt  hep-th/{#1}}}
\def\heplat#1 {{\tt  hep-lat/{#1}}}
\def\adp#1#2#3   
		{{\it Adv.\ Phys.\ }{\bf #1} (#2) #3}
\def\ap#1#2#3    
		{{\it Ann.\ Phys.\ (NY) }{\bf #1} (#2) #3}
\def\app#1#2#3   
		{{\it Astropart.\ Phys.\ }{\bf #1} (#2) #3}
\def\appol#1#2#3 
		{{\it Acta Phys.\ Polon.\ }{\bf #1} (#2) #3}
\def\arnps#1#2#3 
		{{\it Ann.\ Rev.\ Nucl.\ Part.\ Sci.\ }{\bf #1} (#2) #3}
\def\atmp#1#2#3 
		{{\it Adv.\ Theor.\ Math.\ Phys.\ }{\bf #1} (#2) #3}
\def\cpc#1#2#3   
		{{\it Comput.\ Phys.\ Commun.\ }{\bf #1} (#2) #3}
\def\cmp#1#2#3   
		{{\it Comm.\ Math.\ Phys.\ }{\bf #1} (#2) #3}
\def\dmj#1#2#3   
		{{\it Duke Math.\ J. }{\bf #1} (#2) #3}
\def\epjc#1#2#3  
		{{\it Eur.\ Phys.\ J. }{\bf C #1} (#2) #3}
\def\jmp#1#2#3   
		{{\it J.\ Math.\ Phys.\ }{\bf #1} (#2) #3}
\def\jgp#1#2#3   
		{{\it J.\ Geom.\ Phys.\ }{\bf #1} (#2) #3}
\def\jphg#1#2#3   
		{{\it J. Phys.\ }{\bf G #1} (#2) #3}
\def\cqg#1#2#3   
		{{\it Class.\ and Quant.\ Grav.\ }{\bf #1} (#2) #3}
\def\hpa#1#2#3   
		{{\it Helv.\ Phys.\ Acta }{\bf #1} (#2) #3}
\def\jhep#1#2#3 {{\it J. High Energy Phys.\ }{\bf #1} (#2) #3}
\def\lmp#1#2#3   
		{{\it Lett.\ Math.\ Phys.\ }{\bf #1} (#2) #3}
\def\npa#1#2#3   
		{{\it Nucl.\ Phys.\ }{\bf A #1} (#2) #3}
\def\npb#1#2#3    
		{{\it Nucl.\ Phys.\ }{\bf B #1} (#2) #3}
\def\npps#1#2#3  
		{{\it Nucl.\ Phys.\ }{\bf #1} {\it(Proc.\ Suppl.)} (#2) #3}
\def\pla#1#2#3   
		{{\it Phys.\ Lett.\ }{\bf A #1} (#2) #3}
\def\plb#1#2#3   
		{{\it Phys.\ Lett.\ }{\bf B #1} (#2) #3}
\def\ppnp#1#2#3  
		{{\it Prog.\ Part.\ Nucl.\ Phys.\ }{\bf #1} (#2) #3}
\def\pr#1#2#3    
		{{\it Phys.\ Rev.\ }{\bf #1} (#2) #3}
\def\pra#1#2#3   
		{{\it Phys.\ Rev.\ }{\bf A #1} (#2) #3}
\def\prb#1#2#3   
		{{\it Phys.\ Rev.\ }{\bf B #1} (#2) #3}
\def\prc#1#2#3   
		{{\it Phys.\ Rev.\ }{\bf C #1} (#2) #3}
\def\prd #1#2#3  
		{{\it Phys.\ Rev.\ }{\bf D #1} (#2) #3}
\def\pre#1#2#3   
		{{\it Phys.\ Rev.\ }{\bf E #1} (#2) #3}
\def\prep#1#2#3  
		{{\it Phys.\ Rep.\ }{\bf #1} (#2) #3}
\def\prl#1#2#3   
		{{\it Phys.\ Rev.\ Lett.\ }{\bf #1} (#2) #3}
\def\ptp#1#2#3   
		{{\it Prog.\ Theor.\ Phys.\ }{\bf #1} (#2) #3}
\def\rmp#1#2#3   
		{{\it Rev.\ Mod.\ Phys.\ }{\bf #1} (#2) #3}
\def\zpc#1#2#3   
		{{\it Z.\ Physik }{\bf C #1} (#2) #3}
\def\mpl#1#2#3a  
		{{\it Mod.\ Phys.\ Lett.\ }{\bf A #1} (#2) #3}
\def\mplb#1#2#3  
		{{\it Mod.\ Phys.\ Lett.\ }{\bf B #1} (#2) #3}
\def\sjnp#1#2#3  
		{{\it Sov.\ J.\ Nucl.\ Phys.\ }{\bf #1} (#2) #3}
\def\jetp#1#2#3  
		{{\it Sov.\ Phys.\ JETP\/ }{\bf #1} (#2) #3}
\def\jetpl#1#2#3  
		{{\it Sov.\ Phys.\ JETP Lett.\ }{\bf #1} (#2) #3}
\def\zetf#1#2#3  
		{{\it Zh.\ Eksp.\ Teor.\ Fiz.\ }{\bf #1} (#2) #3}
\def\yf#1#2#3    
		{{\it Yad.\ Fiz.\ }{\bf #1} (#2) #3}
\def\nc#1#2#3    
		{{\it Nuovo Cim.\ }{\bf #1} (#2) #3}
\def\joth#1#2#3  
		{{\it J.\ Operator Theory }{\bf #1} (#2) #3}

\def\ijmpa#1#2#3 
		{{\it Int.\ J.\ Mod.\ Phys.\ }{\bf A #1} (#2) #3}
\def\ijmpb#1#2#3 
		{{\it Int.\ J.\ Mod.\ Phys.\ }{\bf B #1} (#2) #3}

\begin{document}

\title{A gauge invariant regulator for the ERG}


  
\author{S. Arnone, Yu.A. Kubyshin\footnote{On leave of absence from
the Institute for Nuclear Physics, Moscow State University, 119899 Moscow,
Russia.} \hspace{-.4em}, \hspace{.05em} 
T.R. Morris and J.F. Tighe}

\address{Department of Physics and Astronomy,\\
University of Southampton\\
Highfield, Southampton SO17 1BJ, U.K.\\
E-mail: sa, kubyshin, trmorris, jft @hep.phys.soton.ac.uk}


\maketitle

\vspace{-6cm}
\begin{flushright}
SHEP 01-07
\end{flushright}
\vspace{6cm}

\abstracts{A gauge invariant regularisation for dealing with pure
Yang-Mills theories within the exact renormalization group approach is
proposed.  
It is based on the regularisation via covariant higher derivatives 
and includes auxiliary Pauli-Villars fields which amounts to a
spontaneously broken  
$SU(N|N)$ super-gauge theory. We demonstrate perturbatively that the extended 
theory is ultra-violet finite in four dimensions  
and argue that it has a sensible limit when the regularization cutoff 
is removed.}


\newcommand{\nn}{\nonumber}

\def\r{\rightarrow}
\def\err{\end{array}}
\def\bp{{\bf p}}
\def\bk{{\bf k}}
\def\bq{{\bf q}}
\def\ttau{\tilde{\tau}}
\def\tchi{\tilde{\chi}}
\def\trho{\tilde{\rho}}
\def\teps{\tilde{\epsilon}}
\def\tnu{\tilde{\nu}}
\def\tgamma{\tilde{\gamma}}


\section{Introduction}

In an earlier work\cite{TM1} one of us proposed a manifestly 
gauge invariant formulation of the exact renormalization group (ERG) 
approach for pure Yang-Mills 
theories.\cite{TM2,TM3,TM4} 
In the same article it 
was suggested to use the gauge invariant regularisation by higher 
covariant derivatives\cite{S1} within this formulation. However, as it 
is known, this regularisation fails to regulate certain one-loop 
ultraviolet divergences.\cite{S1,LZJ} They are regularised by 
introducing additional regulating fields.\cite{S2} 
It was realized that adding bosonic as well as fermionic fields 
to cancel ultraviolet divergences in such a way as 
to maintain the gauge invariance results in a spontaneously broken
$SU(N|N)$ gauge theory.\cite{TM3}  

In the present contribution we start with the $SU(N)$ pure gauge theory in 
$D$ space-time dimensions with the action  
\be
S_{YM} = \frac{1}{2} \int d^{D}x \, \tr \left( F_{\mu \nu} F^{\mu \nu} \right)
\label{S-YM0}
\ee
and extend it so as to include additional fields and covariant higher 
derivatives as regulators. 
The latter introduce also a scale $\Lambda$ which plays the 
r\^{o}le of the effective momentum cutoff. We also add a scalar Higgs field to 
give masses of order $\Lambda$ to some of the 
regulating fields, so that the massive ones behave precisely as
Pauli-Villars fields.
Our aim is to check the consistency 
of such a regularisation scheme. We show that the extended theory 
in four dimensions is indeed 
ultraviolet finite when the cutoff $\Lambda$ is kept finite.
In the continuum limit, $ \Lambda \r \infty$, the massive 
unphysical fields become infinitely heavy and decouple from the theory. 
In addition the unphysical gauge field sector decouples from the 
sector of physical fields in this limit. We argue that in this way the 
initial theory (\ref{S-YM0}) is 
recovered in the continuum limit.  

The plan of the article is as follows. In \secti 2 we describe in more detail 
the main idea of the gauge invariant regularisation scheme. In \secti 3  
the graded Lie algebra of $SU(N|M)$ is discussed and the field content and 
the action of the $SU(N|N)$ gauge theory are presented. In \secti 4 we 
consider some potential problems due to the presence of the Pauli-Villars 
fields using the example of $U(1|1)$ quantum mechanics and discuss 
the decoupling of the unphysical sector. \csec 5 is devoted to 
the proof of finiteness of the regularised theory. 
\csec 6 contains a summary of results. 

\section{Regularised extension of the $SU(N)$ Yang-Mills theory: 
general idea}

As with the case of scalar and fermionic theories, our aim is 
to regularise a pure Yang-Mills theory in a way appropriate 
for the ERG approach, \ie introducing a 
physical cutoff $\Lambda$ which 
sets the physical scale. Another requirement is that the regularisation 
should be manifestly gauge invariant. 

We will adopt the 
point of view that a suitable regularisation for applications within 
the ERG approach must make the theory finite at least in the 
perturbative sense, \ie at each order of 
perturbation theory. 
We believe that this is sufficient to make the theory finite within the 
non-perturbative treatment. Indeed, the perturbative sector is regularised 
by construction. 
As expected, if Feynman diagrams in the perturbative sector are regulated
so as to be free of ultraviolet divergences, then all contributions 
in the expansion around, for example, an instanton, are also finite. 

We implement the regularisation in two steps. 
The first step is to introduce the higher covariant derivative
regularisation.\cite{S1} 
For this we modify the canonical part of the effective 
action and the propagator as follows:
\be
 S= \frac{1}{2} \tr \int d^{D}x \; F_{\mu \nu} \, c^{-1} 
\left( -\frac{\nabla^{2}}{\Lambda^{2}} \right) \cdot 
F^{\mu \nu} + \ldots,   \label{S1}
\ee
where $\nabla_{\mu} = \partial_{\mu} - igA_{\mu}$ and the 
dot means that
the covariant derivative acts by commutation.
Here $c\left( -\nabla^{2}/\Lambda^{2} \right)$ is a (smooth) 
ultraviolet cutoff profile satisfying $c(0) = 1$ so that at low energies 
the propagator is unaltered, and $c(z) \rightarrow 0$ as 
$z \rightarrow \infty$ sufficiently fast so that all Feynman diagrams are 
expected to be ultraviolet regulated. 

As mentioned in the introduction, some one-loop diagrams 
remain unregularised whatever the choice of the regulating function
$c^{-1}$: they require some additional regularisation. 
Such regularisation can be provided by adding some auxiliary
fields.\cite{BS,S2} 
Introducing these fields just formally - in the way it is 
done in many text books\cite{BSh,itz} - breaks gauge invariance. 
To preserve it, 
we introduce them 
by extending the theory (\ref{S1}) 
in a particular way. 
Thus, the second step is to supplement the theory with 
regulating fields. We will see that they include 
\begin{description}
\item[-] an additional bosonic field $A^{2}_{\mu}$; 
\item[-] fermionic (Pauli-Villars) fields $B_{\mu}$, $\bar{B}_{\mu}$;
\item[-] two bosonic (Pauli-Villars) scalar fields $C^{1}$.
\end{description}
The fields $B_{\mu}$, $C^{1}$ and $C^{2}$ 
have masses of order $\Lambda$.

It turns out that this scheme eventually results in the spontaneously broken
$SU(N|N)$ gauge theory.\cite{TM3} In particular, the physical field, \ie 
the gauge field $A_{\mu} \equiv A^{1}_{\mu}$ of the initial $SU(N)$
Yang-Mills theory (\ref{S1}), together with the fields $A^{2}_{\mu}$ and
$B_{\mu}$, $\bar{B}_{\mu}$ form a super-gauge multiplet of $SU(N|N)$, and  
$B_{\mu}$, $C^{1}$ and $C^{2}$ get their masses via the Higgs mechanism.

\section{Graded Lie algebra of $SU(N|M)$ and $SU(N|N)$ gauge theory}

Elements of the graded Lie algebra\cite{Bar} of $SU(N|M)$  
are given by Hermitian $(N+M) \times (N+M)$ matrices 
\be
{\cal H} = \left( \begin{array}{cc}
                   H_{1} & \theta \\
                   \theta^{\dagger} & H_{2}
                   \end{array}
            \right).
\ee
Here $H_{N}$ ($H_{M}$) is an $N \times N$ ($M \times M$) Hermitian matrix with
complex bosonic elements, $\theta$ is an $N \times M$ matrix composed of 
complex Grassmann numbers. ${\cal H}$ is required to be supertraceless 
\be
\str({\cal H}) :=\tr (H_{1}) - \tr (H_{2}) = 0.
\ee
The bosonic sector of the $SU(N|M)$ algebra forms the $SU(N) \times
SU(M) \times U(1)$ subalgebra. 

Let us now specialise to $M=N$, the case we will be interested in. 
The $2N \times 2N$ identity matrix, $\one_{2N}$, satisfies 
$\str \one_{2N} = 0$ and, therefore, is an element of $SU(N|N)$. 
An arbitrary $2N \times 2N$ supermatrix ${\cal X}$ can be written as  
\be
{\cal X} = \frac{1}{2N} \str ({\cal X}) \sigma_{3} + 
\frac{1}{2N} \tr({\cal X}) \one_{2N} + {\cal X}^{A} T_{A}, 
\ee
where 
\be
\sigma_{3} = \left( \begin{array}{cc}
                   \one_{N} & 0 \\
                   0 & -\one_{N}
                   \end{array}
            \right),
\ee
and the other $SU(N|N)$ generators, $T_A$, are made of complex numbers. They can be chosen such that 
$\str(T_{A}) = 0 = \tr(T_{A})$. 
The index $A$ runs over $2(N^{2}-1)$ bosonic and $2N^{2}$ fermionic 
indices. For ${\cal H} \in SU(N|N)$ 
\be
{\cal H} = {\cal H}^{0} \one_{2N} + {\cal H}^{A}T_{A}. 
\ee
We also define the Killing super-metric in the $T_A$ subspace 
\be
g_{AB} = \frac{1}{2} \str\left( T_{A} T_{B} \right). 
\ee
$g_{AB}$ is symmetric when both indices $A$ and $B$ are bosonic, antisymmetric 
when both are fermionic and is zero when one is bosonic and another is 
fermionic. 

Let us turn to the construction of the 
$SU(N|N)$ extension of the regularised $SU(N)$ 
gauge theory (\ref{S1}). A basic ingredient is the super-gauge field 
${\cal A}_{\mu}$ which takes values in the graded Lie algebra $SU(N|N)$ and 
therefore can be written as  
\bea
{\cal A}_{\mu} & = & {\cal A}^{0}_{\mu} \one_{2N} + \tilde{\cal A}_{\mu}, \nonumber \\
\tilde{\cal A}_{\mu} & = & \left( \begin{array}{cc}
                   A^{1}_{\mu} & B_{\mu} \\
                   \bar{B}_{\mu} & A^{2}_{\mu}
                   \end{array}
            \right) = {\cal A}^{A}_{\mu} T_{A}. 
\eea
Here $A^{1}_{\mu} \equiv A_{\mu}$ is the physical $SU(N)_{1}$ gauge field, 
\ie the gauge field in the initial theory (\ref{S-YM0}), while the
fields $A^{2}_{\mu}$ and $B_{\mu}$ are part of the regulating structure. 
It will turn out later that the field 
${\cal A}^{0}_{\mu}$ is not dynamical and being a $U(1)$ factor does not
couple to the other fields. Therefore in the rest of the 
article we will consider the $\tilde{\cal A}_{\mu}$ field only. 
We will also omit the tilde to ease notation.  

The action of the super-gauge field is taken to be 
\be
{\cal S}_{YM} = \frac{1}{2} {\cal F}_{\mu \nu} \{ c^{-1} \} {\cal F}^{\mu \nu}, 
\label{S-YM}
\ee
where the gauge field strength is defined as ${\cal F}_{\mu \nu} =
\partial_{\mu}{\cal A}_{\nu} - \partial_{\nu}{\cal A}_{\mu} -ig[{\cal
A}_{\mu},{\cal A}_{\nu}]$. 
For a given kernel $W$ and two super-matrices $u(x)$ and $v(x)$ the wine 
$u\{ W \} v$ is defined as
\be
u \{W\}v := \str \int d^{D}x \, u(x) 
W \left( -\frac{\nabla^{2}}{\Lambda^{2}}\right) \cdot v(x).   
\ee
The regulating function $c^{-1}$ is chosen to be a polynomial in 
$(-\nabla^{2}/\Lambda^{2})$ of rank $r$. 

Next we introduce a super-scalar field
\be
{\cal C} = \left( \begin{array}{cc}
                   C^{1} & D \\
                   \bar{D} & C^{2}
                   \end{array}
            \right),   \label{C-def}
\ee
with the action 
\be
{\cal S}_{\cal C} = \nabla_{\mu} \cdot{\cal C} \{ \tilde{c}^{-1}\}
\nabla^{\mu} \cdot {\cal C} + 
\frac{\lambda}{4} \str \int d^{D}x \; \left( {\cal C}^{2} - \Lambda^{2}
\right)^{2},          \label{S-S}
\ee
where $\tilde{c}^{-1}$ is another regulating function. It is assumed 
to be a polynomial in $(-\nabla^{2}/\Lambda^{2})$ of rank $\rtil$. 
The field (\ref{C-def}) is not assumed supertraceless. 
It acquires a vacuum expectation value which may be taken to be 
$<{\cal C}> = \Lambda \sigma_{3}$.  
Shifting ${\cal C} \rightarrow \Lambda \sigma_{3} + {\cal C}$, the  
action in the scalar sector becomes  
\bea
{\cal S}_{\cal C} & = & -g^{2} \Lambda^{2} [{\cal A}_{\mu},\sigma_{3}] \{
\tilde{c}^{-1}\} [{\cal A}_{\mu},\sigma_{3}] 
 - 2ig \Lambda [{\cal
A}_{\mu},\sigma_{3}]  \{ \tilde{c}^{-1}\} \nabla^{\mu} \cdot {\cal C} \nonumber \\
& + & \nabla_{\mu} \cdot {\cal C}\{ \tilde{c}^{-1}\} \nabla^{\mu} \cdot {\cal C} 
+ \frac{\lambda}{4} \; \str \int\! d^D x \, \left( \Lambda \{ \sigma_{3},{\cal C}\}_{+} + 
{\cal C}^{2} \right)^{2}.   
\eea
One can check that the fields $B_{\mu}$, $C^1$ and $C^{2}$ acquire masses 
of order $\Lambda$, thus behaving precisely as Pauli-Villars fields.  

It is convenient to impose 't Hooft's gauge fixing condition 
\be
\partial_{\mu} {\cal A}^{\mu} +ig \frac{\Lambda}{\xi}\tilde{c}^{-1} \hat{c}
[\sigma_{3},{\cal C}] = 0,  
\ee
where $\hat{c}^{-1}(-\partial^{2}/\Lambda^{2})$ is a polynomial of rank $\rhat$. 
Note that here the regulating function is not covariantised. 
The corresponding gauge fixing term in the action is  
\bea
{\cal S}_{GF} & = & \xi \partial_{\mu} {\cal A}^{\mu} \cdot \hat{c}^{-1} \cdot 
\partial_{\nu} {\cal A}^{\nu} + 2ig \Lambda (\partial_{\mu} {\cal A}^{\mu}) 
\cdot \tilde{c}^{-1} \cdot [\sigma_{3},{\cal C}] \nonumber \\
& - & g^{2} \frac{\Lambda^{2}}{\xi} [\sigma_{3},{\cal C}] \cdot 
\tilde{c}^{-2} \hat{c} \cdot [\sigma_{3},{\cal C}], 
\eea
where $f \cdot W \cdot g \doteq \str \int \!\!\!\int d^D x d^D y \, f(x) W_{xy}
g(y)$ and $W_{xy}$ is the inverse Fourier transform of the kernel $W \left(
\frac{p^2}{\Lambda^2} \right)$. 
The Faddeev-Popov ghost super-fields form the $SU(N|N)$ supermatrix 
\be
\eta = \left( \begin{array}{cc}
                   \eta^{1} & \phi \\
                   \psi & \eta^{2}
                   \end{array}
            \right). 
\ee
The action of the ghost sector is given by 
\bea
{\cal S}_{ghost} & = & - \bar{\eta} \cdot \hat{c}^{-1}\tilde{c} 
\cdot \partial_{\mu} \nabla^{\mu} \cdot \eta  \nonumber \\
& - & \int d^{D}x \; \str\left\{ \frac{\Lambda}{\xi} [\sigma_{3}, \bar{\eta}] 
\left( \Lambda [\sigma_{3},\eta] + [{\cal C},\eta] \right) \right\}.
\eea

In order to keep the high momentum behaviour of the $\A$ propagator
unchanged by the introduction of the $\C$ field and gauge fixing, we
require the  ranks of our polynomial cutoff
functions to be bounded as 
\be \label{behav}
\rhat \geq r \geq \rtil.
\ee

\section{Potential problems in the unphysical sector}

The quadratic part of the action (\ref{S-YM}) is equal to 
\bea
{\cal S}_{YM} & = & \int d^{D}x \frac{1}{2} \left[ \tr (F^{1}_{\mu \nu})^{2} - 
\tr (F^{2}_{\mu \nu})^{2}   \right. \nonumber \\ 
& - & \left. 2\tr 
\left(\partial_{\mu} \bar{B}_{\nu} - \partial_{\nu} \bar{B}_{\mu} \right)
\left( \partial_{\mu} B_{\nu} - \partial_{\nu} B_{\mu} \right) + \ldots \right].
\nonumber
\eea
The appearance of the term with negative sign could potentially be a problem
due to an instability in the theory. 
Below we consider the example of $U(1|1)$ quantum mechanics which shows   
that it is rather a sign of the loss of unitarity.

Let us define the Hermitian super-position as 
\be
{\cal X} = \left( \begin{array}{cc}
                   x_{1} & \psi \\
                   \bar{\psi} & x_{2}
                   \end{array}
            \right),  
\ee
and consider the model with a simple harmonic potential. The Lagrangian is 
given by 
\be
L = \frac{1}{2} \str\dot{\cal X}^{2} - \frac{1}{2} \str{\cal X}^{2}.
\ee
The conjugate momentum variables are equal to 
\bea
& & p_{x_{1}} = \dot{x}_{1}, \; \; p_{x_{2}} = - \dot{x}_{2}, \; \;
[x_{j},p_{x_{j}}] = i, \nonumber \\
& & p_{\psi} = \dot{\bar{\psi}}, \; \; p_{\bar{\psi}} = - \dot{\psi}. 
\eea
Next we define the $a_{j}$, $a_{j}^{\dagger}$ operators $(j=1,2)$ 
\be
a_{j} = \frac{1}{\sqrt{2}} \left( x_{j} + ip_{x_{j}} \right), \; \; \; 
a_{j}^{\dagger} = \frac{1}{\sqrt{2}} \left( x_{j} - ip_{x_{j}}
\right), 
\ee
with the commutation relations $[a_{i},a_{j}^{\dagger}]=\delta_{ij}$. 
In terms of these operators the Hamiltonian reads
\be
H=\frac{1}{2} (a_{1}^{\dagger} a_{1} + a_{1}a_{1}^{\dagger}) - 
\frac{1}{2} (a_{2}^{\dagger} a_{2} + a_{2}a_{2}^{\dagger}) + \mbox{(fermionic
part)}.
\ee
Now we introduce the vacuum states $|0>_{1}$, $|0>_{2}$ satisfying the 
relations 
\be
a_{1}|0>_{1} = 0, \; \; a_{2}^{\dagger}|0>_{2} = 0,
\ee
and build up sets of $n$-particle states according to the formulae 
\bea
& & |n>_{1} = \frac{1}{\sqrt{n!}} (a_{1}^{\dagger}
)^{n}|0>_{1},  \nonumber \\
& & |n>_{2} = \frac{1}{\sqrt{n!}}
(a_{2} )^{n}|0>_{2}.  
\eea
Note that $a_{2}$ plays the r\^{o}le of the creation operator of the particle
of the second type. 
With such definitions the Hamiltonian of the system is bounded from below. 
In particular,  
\be
H|n>_{2} = + n|n>_{2}.
\ee
Furthermore, it can be shown that these definitions ensure that the vacuum
preserves the $U(1|1)$ symmetry.
However the $n$-particle states in the `2'-sector with odd $n$ possess 
negative norms:  
\be
_{2}<n|n>_{2} = \frac{1}{n!} {_{2}<0|}(a_{2}^{\dagger})^{n}
(a_{2})^{n}|0>_{2} = (-1)^{n} {_{2}<0|0>_{2}}.
\ee
This can be referred to as violation of unitarity (negative probability). 

We would like to mention that the appearance of negative norm states 
as a consequence of a wrong sign in part of the action is not that 
unusual. The Gupta-Bleuler quantization procedure\cite{BSh}
 relies on a modification of the Lagrangian which 
results in a wrong sign appearing in the $A^{0}$ part of the
action. Nonetheless, in the present case there is no analogue of the
Gupta-Bleuler condition.

At finite $\Lambda$, transitions between the $A^{1}_{\mu}$- 
and the $A^{2}_{\mu}$-sector via (massive) Pauli-Villars fields 
are not forbidden - they are just suppressed, thus leading to 
violation of unitarity.  
In the continuum limit ($\Lambda \rightarrow \infty$), however, the 
amplitudes for such transitions vanish and the physical sector decouples from 
the unphysical one.   

To see an example let us consider $D=4$. 
The lowest order $A^{1} A^{2}$ amplitude comes from the term  
\be
\str({\cal A} {\cal A}) \str({\cal A} {\cal A}) \times (\mbox{IR and UV finite
coefficient})
\ee
in the effective action. 
Gauge invariance and dimensional considerations imply that 
this term is in fact  
\be
\sim \int d^{4}x \frac{1}{\Lambda^{4}} \str({\cal F}{\cal F}) \str({\cal
F}{\cal F}).
\ee
Therefore, it vanishes as $\Lambda \rightarrow \infty$.

\section{Counting of ultraviolet divergences}

Using standard rules for calculating the superficial degree of
divergence\cite{itz} of a one-particle-irreducible (1PI) diagram in $D$ 
space-time dimensions, we get
\be \label{sdd1}
\begin{array}{l}
{\ds {\cal D}_{\Gamma} = D L - (2r+2) \, I_{\A} -(2\rtil+2) \, I_{\C}
-(2 \rhat -2\rtil+2) \, I_{\eta}  + \sum_{i=3}^{2r+4} (2r+4-i) \,
V_{{\A}^i} }\nonumber\\ 
{\ds + \sum_{j=2}^{2\rtil+2} (2\rtil+2-j) \, V_{{\A}^j {\C}}
+ \sum_{k=1}^{2\rtil+2} (2\rtil+2-k) \,
V_{{\A}^k {\C}^2} + (2 \rhat -2\rtil+1) \, V_{{\eta}^2 \A}, }
\end{array}
\ee
where $L$ is the number of loops and $I_{f}$ and $V_{f}$ correspond to the
number of internal lines and vertices of $f$-type
respectively. In \eq{sdd1}, inequalities (\ref{behav}) have already been 
assumed for 
the degree of divergence of the vector propagator to be counted properly. 

As it stands, \eq{sdd1} does not account properly for 1PI diagrams with
external anti-ghost lines. In fact, the whole momentum dependence of the
$V_{{\eta}^2 \A}$ vertex 
is counted as flowing into the loop, without taking into account the 
fact that such a
dependence is actually only carried by $\bar{\eta}$ lines and, thus, that one has to
check whether such lines are external or not. This results in a systematic
overestimate of \dga. In order to 
remedy this and, thus, improve our upper bound, \dga, we add 
$-(2 \rhat-2\rtil+1) E_{\bar{\eta}}^{\A}$, with $E_{\bar{\eta}}^{\A}$ 
being the number
of external anti-ghost lines which enter $V_{{\eta}^2 \A}$ vertices.\\
Therefore, the 
improved formula for the superficial degree of divergence
is   
\be \label{sdd1im}
\begin{array}{l}
{\ds {\cal D}_{\Gamma} = D L - (2r+2) \, I_{\A} -(2\rtil+2) \, I_{\C}
-(2\rhat-2\rtil+2) \, I_{\eta}  + \sum_i (2r+4-i) \, V_{{\A}^i} +}\nonumber\\
{\ds \sum_j (2\rtil+2-j) \, V_{{\A}^j {\C}}
+ \sum_{k} (2\rtil+2-k) \,
V_{{\A}^k {\C}^2} + (2\rhat-2\rtil+1) \left( V_{{\eta}^2 \A} - E_{\bar{\eta}}^{\A} \right). }
\end{array}
\ee 
The variables upon which ${\cal D}_{\Gamma}$ is dependent can be easily
related to the number of external lines of each type, $E_{f}$, as
\be \label{usefulrel}
\begin{array}{l}
{\ds L = 1 + I_{\A} + I_{\C} + I_{\eta} - \sum_i
V_{{\A}^i} - \sum_j V_{{\A}^j {\C}}
- \sum_{k} V_{{\A}^k {\C}^2} -
V_{{\eta}^2 \A} - V_{{\eta}^2 \C} - V_{{\C}^3} - V_{{\C}^4}, }\nonumber\\
{\ds E_{\A} = -2 I_{\A} +  \sum_i
i V_{{\A}^i} + \sum_j j V_{{\A}^j {\C}} +
\sum_{k} k V_{{\A}^k {\C}^2} +
V_{{\eta}^2 \A}, }\nonumber\\
{\ds E_{\C} = -2 I_{\C} + \sum_j V_{{\A}^j {\C}} +
2 \, \sum_{k} V_{{\A}^k {\C}^2} +
 3 V_{{\C}^3} + 4 V_{{\C}^4} + V_{{\eta}^2 \C}, }\nonumber\\
E_{\eta} = {\ds E_{{\eta}}^{\A} + E_{{\eta}}^{\C} +
E_{\bar{\eta}}^{\A} + E_{\bar{\eta}}^{\C} = -2 I_{\eta} + 2 V_{{\eta}^2 \A}
+ 2 V_{{\eta}^2 \C}.} 
\end{array}
\ee
In the last of the above relations, to ensure consistency with previous
notation we split external ghost lines
according to the vertices they are attached to. Thus
$E_{(\bar{\eta})\eta}^{f}$, $f \!=\! \A,\C$, is the number of external (anti-) ghost lines
entering $V_{{\eta}^2 f}$ vertices; they satisfy  
the expected constraint $E_{{\eta}}^{\A} + E_{{\eta}}^{\C} =
E_{\bar{\eta}}^{\A} + E_{\bar{\eta}}^{\C}$.
The first of \eq{usefulrel} is valid for connected diagrams only, as 
the first term in the r.h.s. - representing the number of connected
components - has been set to $1$. 

By making use of 
the above formulae, it is possible to rewrite
${\cal D}_{\Gamma}$ in a more useful form, 
independent of internal lines,
\be \label{sdd2}
\begin{array}{l}
{\ds {\cal D}_{\Gamma} = (D-2r-4) \, (L-2) - E_{\A} -(r-\rtil+1) \,
E_{\C} -2 (r+\rtil-\rhat +1) \,
E_{\bar{\eta}}^{\C} - \phantom{\sum_{j=1}^2} } \nonumber\\
{\ds (2r+3) E_{\bar{\eta}}^{\A} - (r -\rtil + 1) \sum_j
V_{{\A}^j {\C}} + (r-3\rtil-1) \, V_{{\C}^3}  + 2 (r-2\rtil) \,
V_{{\C}^4} }\nonumber\\
{\ds + (r+\rtil-2\rhat -1) \, V_{{\eta}^2 \C} + 2 (D-r-2). }
\end{array}
\ee  
    
As stated by the convergence theorem,\cite{itz} if the superficial degree of
divergence of all the connected proper sub-diagrams of a given diagram $G$
is negative, then the Feynman integral corresponding to $G$ is absolutely
convergent. Therefore, we have to find constraints (if they
exist) over $r$, $\rtil$ and $\rhat$, such that ${\cal D}_{\Gamma}$ is
negative for all the possible diagrams at any loop order. 

However, not all the diagrams are regularised this way.
For example, the degree of divergence of the one-loop diagrams with just
external $\A$ lines is 
$D - E_{\A}$, \ie independent of the parameters $r$, $\rtil$ and $\rhat$
and hence no conditions can be found for it to be negative.
We will start with the analysis of diagrams with two or more loops in an 
arbitrary number of dimensions and, after, we will return to one-loop
diagrams 
to show the finiteness of the theory in four dimensions only.

\subsection{Multiloop graph analysis}

In order for 
every possible 1PI diagram to have a negative \dga, 
we can impose all coefficients in \eq{sdd2} to be negative and, thus, get
sufficient conditions.\\  
This amounts to the following relations
\be \label{conditions}
r>D-2, \quad \qquad r<2\rtil , \quad \qquad \rhat<r+\rtil+1,
\ee
together with \eq{behav}. It is easy to see that there are integers $r$,
$\rtil$, $\rhat$ satisfying eqs (\ref{behav}), (\ref{conditions}).
To get proper bounds, $r$, $\rtil$, $\rhat$ should be considered as real
numbers, the restriction to integers being taken at the end. As a matter of
fact it is consistent to take these parameters real having in mind more
general cutoff functions (analytic around the origin, $p=0$, and with
asymptotic behaviour $c^{-1} \sim
{p^{2r} \over \Lambda^{2r}}$ etc.).

The conditions (\ref{conditions}) imply a lower bound on $\rtil$, 
$\rtil>{D \over 2} -1$, as well.   
Some of the relations (\ref{conditions}) - those setting $D$-dependent
lower bounds 
on $r$ and $\rtil$ - may be expected to be also necessary, as the higher
the space-time dimension, the more divergent the diagrams.

However, physics does not provide any reasonable arguments to 
explain {\emph{upper}} bounds on $\rhat$ and $r$, apart from $r \leq \rhat$
(cf \eq{behav}). 
In fact they are not necessary, as we now show.

Let us denote by $\S$
the collection of triples $(r,\rtil,\rhat)$ such that \dga \, is negative for 
any given set of 1PI diagrams and \eq{behav} holds.\\[0.3cm] 
{\bf Proposition}: {\sl 
If $(r_0,\rtilo,\rhato) \in \S$, then the subset 
$\big\{ (r, \rtil, \rhat) \;\, s.t. \,\; r \!\geq \! r_0, \, \rtil \! = \!
\rtilo, \, \rhat \! \geq \!
\rhato,\\ 0 \! < \! \rtilo \! \leq \! r \! \leq \! \rhat \big\} \subset
\S$. }\\[0.3cm] 
\newpage
{\sl Proof:}\\[0.3cm] 
The proof is essentially based on the one-particle-irreducibility of 
diagrams.

The whole dependence of \eq{sdd2} on $\rhat$ amounts to $ 2 \rhat \, \big( E_{\bar{\eta}}^{\C} -
V_{{\eta}^2 \C} \big)$, which is always non-positive as it is not possible to
have more external anti-ghost lines entering $V_{{\eta}^2 \C}$ vertices
than $V_{{\eta}^2 \C}$ vertices themselves. Thus, increasing $\rhat$ above
$\rhato$ can only
decrease an already negative \dga.

As far as $r$ is concerned, it enters \eq{sdd2} as 
\be
r \Big( -2L + 2 - E_{\C} -2 
E_{\bar{\eta}}^{\C} - 2 E_{\bar{\eta}}^{\A} - \sum_j
V_{{\A}^j {\C}} + V_{{\C}^3}  + 2 \, V_{{\C}^4} + V_{{\eta}^2 \C} \Big) = 2
r \Big( \sum_i V_{{\A}^i} - I_{\A} \Big),
\ee
where the last equality follows by using \eq{usefulrel} or directly
from \eq{sdd1im}.\\
This contribution is always non-positive as we know that in a 1PI diagram every 
$V_{{\A}^i}$ vertex must attach to at least two internal $\A$ lines.
Therefore increasing $r$ above $r_0$ can only cause \dga \, to decrease
further. \hfill $\Box{4.7pt}$ 

Using the above proposition, we see that the second and the third
inequalities in \eq{conditions} are not necessary, 
and we are thus left with the sufficient relations $r>D-2$, $\rtil>{D \over 2}
-1$ and $\rhat \geq r \geq \rtil >0$. 
For the case that the inverse cutoff functions are polynomials these
inequalities imply
\be \label{integers}
r \geq D-1, \qquad \qquad \rtil \geq \left[ \frac{D-2}{2} \right] +1,
\ee
$[x]$ being the integer part of $x$.
The above conditions are also necessary, as they ensure finiteness in the 
two two-loop vacuum diagrams with only $\A^3$ and $\C^4$ vertices respectively. 

\subsection{One loop diagram analysis}

\begin{figure}[!h]
\begin{center}
\epsfxsize=25pc 
\epsfbox{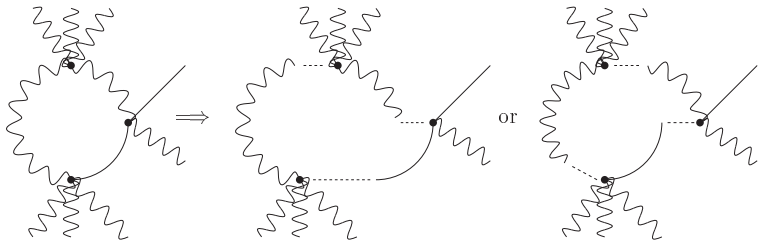} 
\caption{Tadpole-like pieces.  \label{fig:tadpole}}
\end{center}
\end{figure}

To perform the analysis of one-loop diagrams we adopt the strategy of {\it
divide and conquer:} we cut them up 
into tadpole-like pieces, defined as the sub-diagrams which contain just 
one internal
propagator attached to one vertex.\cite{TM3} This can be done in two 
different ways, according to  
which propagator remains attached to the vertex being cut (see
fig. \ref{fig:tadpole} for an example).\\

We then compute the degree of divergence of every possible piece
we can end up with, aiming to show that 
they all contribute negatively to the overall \dga. If this is the case - and 
it is indeed - we are just left with the analysis of the simplest possible
one-loop graphs, as any other can be obtained by adding tadpole-like pieces,
which causes \dga \ to decrease further. 

In other words, 
we can
always bound from above the degree of divergence of a one-loop diagram by
removing tadpole-like pieces one by one and joining together the rest of
the diagram - hence increasing the overall
\dga. Eventually we will be left with a very simple graph, usually a proper
tadpole. 

The first part of the analysis, that is calculating \dga \ for such 
``components'', is straightforward: by inspection of
\eq{sdd1} it is easy to appreciate that all the possible sub-diagrams
contribute negative terms within the bounds we have already set on
$r$, $\rtil$ and $\rhat$.
Attaching an $\A$ propagator to a $V_{\eta^2 \A}$ vertex is the only case
which needs some comment. In this case 
 \dga $= 2\rhat -2r-2\rtil-1$ 
 can be positive if $\rhat$ is large enough.
However, it is not possible to add just a single $V_{\eta^2 \A}$ vertex,
since it would require the introduction of just a single external ghost
line, which is forbidden.
Adding two such vertices make such a contribution convergent as seen in the 
correction introduced in \eq{sdd1im}.

\begin{figure}[h]
\begin{center}
\epsfxsize=25pc 
\epsfbox{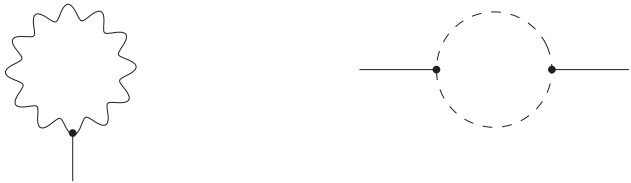} 
\caption{Graphs related to the constraints, eq.(\ref{furcons}).  \label{fig:furcons}}
\end{center}
\end{figure}

The second part of the analysis, \ie showing that all the simplest possible 
one-loop diagrams\footnote{except those whose \dga \ is independent of the
ranks of the cutoff functions, \eg graphs with no external $\C$ or $\eta$
lines and with up to $D$ external $\A$ legs.} can be 
regulated by a suitable choice of $r$, $\rtil$ and $\rhat$, is quite long but
straightforward as well.  
We find two further constraints on the ranks of the cutoff functions
\be \label{furcons}
r-\rtil >{D \over 2} -1, \qquad \qquad \rhat-\rtil >{D \over 4} -1,
\ee
which come from the graphs sketched in fig. \ref{fig:furcons}.

In order to deal with the diagrams that remain unregularised despite the
introduction of (covariantised) cutoff functions, we now specialise to $D=4$.
Those graphs, with no external $\C$ or $\eta$
lines and with up to $4$ external $\A$ legs, are finite
due to the cancellation of the ultraviolet 
divergences between the contributions of the 
bosonic and fermionic propagators corresponding to internal lines. 
This cancellation will be referred to as the supertrace mechanism. 

To illustrate it, let us sketch the calculation of the one-loop 
diagram with two external ${\cal A}$-lines and two internal ${\cal A}$-lines. 
The terms of the perturbation theory expansion which generate this type of 
diagram, schematically omitting the Lorentz indices\footnote{Beware that
the commutators do not vanish once these are taken into account!}
involve the product of two vertices:
\be
\str \left( [{\cal A}(x),{\cal A}(x)]{\cal A}(x) \right) 
\str \left( {\cal A}(y)[{\cal A}(y),{\cal A}(y)] \right),  \label{diag}
\ee
where ${\cal A}$ stands for the gauge field or its derivative. 
The leading part of the 
propagator between the ${\cal A}^{A}$ and ${\cal A}^{B}$ fields 
in the momentum representation is proportional to $g^{AB}$. Using the 
completeness relation for the generators $T_{A}$ it is easy to show that 
by Wick pairing 
\be
\str ({\cal X}{\cal A}(x)) \str ({\cal A}(y){\cal Y})\wick{24}{14}
= \left[ \frac{1}{2}\str ({\cal X}{\cal Y})  
- \frac{1}{4N} ( \tr  {\cal X} \;\str {\cal Y}
+ \str {\cal X} \; \tr  {\cal Y}) \right] \times \Delta (x-y),  
\ee
where $\Delta (x-y)$ is a space-time dependent 
factor coming from the propagator. Applying 
this formula to \eq{diag} one can see that after the first pairing 
it reduces to
\be
\frac{1}{2} 
\str \left([{\cal A}(x),{\cal A}(x)] [{\cal A}(y),{\cal A}(y)] \right) 
 \Delta (x-y). \label{diag1}
\ee 
Here we have used the cyclicity property of the supertrace, $\str ({\cal
X}{\cal Y}) = \str ({\cal Y}{\cal X})$, which implies that $\str ([{\cal
X},{\cal Y}])=0$.  
For the next step we use the identity 
\be
\str \left({\cal X} T_{A} {\cal Y} T^{A} \right) = 
\frac{1}{2}\str ({\cal X})\str({\cal Y}) 
- \frac{1}{4N} \left[\str ({\cal X} \sigma_{3}{\cal Y}) + 
str ({\cal X}{\cal Y} \sigma_{3}) \right], \label{splitid}
\ee 
valid for any super-matrices ${\cal X}$ and ${\cal Y}$, which
follows from the already mentioned completeness relation. Using this 
identity we calculate the second $\A(x)$-$\A(y)$ pairing in 
\eq{diag1} and find
that the $\sigma_3$ terms appearing in \eq{splitid} all cancel,
as they must -- to preserve the $SU(N|N)$ invariance, leaving only terms
of the form $\str{\cal A}\,\str{\cal A}$ or $\str{\cal A}\,\str\one$, 
both of which vanish because $\str {\cal A}=\str\one=0$.
This is a demonstration of the supertrace mechanism at work. 

One can check by direct calculation that the supertrace mechanism ensures 
the finiteness of all the diagrams with two and three external ${\cal
A}$-lines.  
For the diagrams with 4 external ${\cal A}$-lines the supertrace mechanism
is not sufficient (at finite $N$). However, these are already finite.
This follows because gauge invariant
effective vertices containing less than four ${\cal A}$s have already been
shown to be finite but gauge invariant effective vertices 
with a {\sl minimum} of four 
${\cal A}$s are already finite by power counting 
and the Ward identities for the $SU(N|N)$ gauge theory.\cite{AKMT}

To summarise, in order to enforce finiteness on all 
diagrams in four dimensions we need to impose the following constraints (cf
\eq{integers} )
\be \label{nms-rel}
r \geq 3, \quad \rtil \geq 2, \quad r-\rtil >1, \quad
\rhat -\rtil >0, \quad \rhat \geq r \geq \rtil >0.
\ee

\section{Summary and conclusions} 

We have analyzed the $SU(N|N)$ gauge theory with the higher covariant
derivative regulators $c \left( -\nabla^{2}/\Lambda^{2}\right)$ and Higgs
field, viewed as a regularised version of the $SU(N)$ Yang-Mills theory. 
Its structure is determined by the requirement that 
it can be used within a manifestly 
gauge invariant formulation of the ERG.\cite{TM1,TM2,TM3} 

The extension includes the physical Yang-Mills field $A_{\mu}^{1} \equiv 
A_{\mu}$ of the initial theory and the regulating fields: 
the bosonic gauge field $A_{\mu}^{2}$, the fermionic Pauli-Villars field 
$B_{\mu}$ and the scalar Pauli-Villars fields $C^{i}$. 
All the regulator fields except $A_{\mu}^{2}$ acquire masses 
proportional to the momentum cutoff $\Lambda$ via the Higgs mechanism. 
The presence of the unphysical regulator fields lead to a source of
unitarity violation in the theory with finite cutoff. 
However, when the regularisation is removed, \ie in the limit 
$\Lambda \rightarrow \infty$, the massive fields $B_{\mu}$ and
$C^{j}$ become infinitely heavy and decouple. As a consequence the physical 
sector, which is the original $SU(N)$ Yang-Mills theory, 
becomes decoupled from the unphysical sector of 
the field $A_{\mu}^{2}$. In this way the unitarity of the 
theory is restored in the continuum limit, \ie when 
$\Lambda \rightarrow \infty$.

We showed that in four dimensions the 
one-loop one-particle-irreducible diagrams with two, three or four external 
${\cal A}$-lines are finite due to the 
peculiar structure of the $SU(N|N)$ supergauge group. The rest of   
the one-loop one-particle-irreducible diagrams and 
all one-particle-irreducible 
diagrams with the number of loops $L \geq 2$ can be made finite by the proper 
choice of the regulating functions $c^{-1}$, $\tilde{c}^{-1}$ and 
$\hat{c}^{-1}$. The necessary and sufficient conditions on the parameters 
of these functions are given in eqs (\ref{nms-rel}). While we only
demonstrate finiteness in four dimensions, we are confident it can be
extended up to eight dimensions. At this point the Ward identities
regarding the 1PI function with four external $\A$-legs no longer guarantee
the finiteness of such contribution.

We expect the use of the $SU(N|N)$ regularised extension 
of the $SU(N)$ pure Yang-Mills theory to open up new possibilities 
of non-perturbative and gauge-invariant 
treatment of Yang-Mills theories in the framework of the ERG approach.

\section*{Acknowledgements}
We would like to thank J.I. Latorre and C. Wetterich for
discussions and valuable comments. Yu.K. acknowledges financial
support from the PPARC grant PPA/V/S/1998/00907 and from 
the Programme "Universities of Russia" 
(grant 990588). T.R.M. and
S.A. acknowledge support from PPARC SPG PPA/G/S/1998/00527. J.F.T. thanks
PPARC for support through a studentship.

\end{document}